\documentclass[usenatbib]{mn2e}
\usepackage{natbibmnfix,graphicx,times}
\usepackage{epstopdf}
\usepackage{amsmath,amssymb}
\newcommand{\apj}{ApJ}

\newcommand{\apjs}{ApJS}
\newcommand{\aap}{A \& A}
\newcommand{\aj}{AJ}
\newcommand{\mnras}{MNRAS}
\newcommand{\physrep}{Physics Reports}
\newcommand{\prd}{Phys Rev D}
\newcommand{\nat}{Nature}

\newcommand{\kmis}{\,\rm{km\,s^{-1}}}

\title[Galaxy surveys, inhomogeneous reionization, and dark energy]{Galaxy surveys, inhomogeneous reionization, and dark energy}

\author[Pritchard, Furlanetto, and Kamionkowski]{Jonathan R. Pritchard$^1$\thanks{Email: jp@tapir.caltech.edu}, Steven R. Furlanetto$^2$\thanks{Email: steven.furlanetto@yale.edu}, Marc Kamionkowski$^1$\thanks{Email: kamion@tapir.caltech.edu} \\
$^1$ California Institute of Technology, Mail Code 130-33, Pasadena, CA 91125, USA\\
$^2$Yale Center for Astronomy and Astrophysics, Yale University, New Haven, CT 06520-8121, USA} 

\begin{document}

\maketitle

\begin{abstract}
We examine the effect of inhomogeneous reionization on the galaxy power spectrum and the consequences for probing dark energy.  To model feedback during reionization, we apply an {\em ansatz} setting the galaxy overdensity proportional to the underlying ionization field.  Thus, inhomogeneous reionization may leave an imprint in the galaxy power spectrum.  We evolve this imprint to low redshift and use the Fisher-matrix formalism to assess the effect on parameter estimation. We show that a combination of low- ($z=0.3$) and high- ($z=3$) redshift galaxy surveys can constrain the size of cosmological HII regions during reionization.  This imprint can also cause confusion when using baryon oscillations or other features of the galaxy power spectrum to probe the dark energy. We show that when bubbles are large, and hence detectable, our ability to constrain $w$ can be degraded by up to $50\%$.  When bubbles are small, the imprint has little or no effect on measuring dark-energy parameters.
 \end{abstract}
 

\section{Introduction} 
\label{sec:intro}

During the epoch of reionization, groups of star-forming regions generate significant numbers of ionizing photons, which may lead to HII regions many Mpc in size \citep*{fzh2004}.  These ionized bubbles grow as further structure forms, eventually merging and causing full reionization of the intergalactic medium (IGM).  Conditions within these HII regions may be significantly different than in the surrounding neutral IGM.  For example, the temperature in these HII regions will be raised by photoionization heating, which is known to suppress star formation in low-mass haloes \citep{rees1986,efstathiou1992,thoul1996,kitayama2000,dijkstra2004}.  Also, the ionizing flux generates more free electrons, which affects the abundance of molecular hydrogen \citep{oh2002}, an important coolant.  These, and other feedback mechanisms, will affect the fraction of baryons that condense in haloes, and in turn modify the number density of directly observable galaxies \citep{bl2001}.  This suppression will be inherently inhomogeneous, as highly biased regions will ionize first \citep{babich2005}.  Understanding the detailed effects of feedback is one of the major remaining challenges in understanding galaxy formation.

Galaxies formed during reionization will be low in mass and faint by comparison to those from later generations of galaxy formation.  This, along with absorption along the line of sight by the IGM, makes it difficult for existing telescope facilities to detect large numbers of early galaxies directly. However, a number of large galaxy surveys now exist, which probe the distribution of galaxies in the lower-redshift regime \citep{efstathiou2002,seljak2005}.  These surveys focus on high-mass luminous objects -- e.g., luminous red galaxies (LRGs) in SDSS \citep{eisenstein2005} -- which are easily detected.  The abundance of such objects will depend in a non-trivial way upon the number of low-mass progenitors, especially upon the amount of condensed gas available for mergers.  Thus, these late-forming galaxies will, indirectly, be affected by the efficiency of galaxy formation during reionization.  Motivated by these arguments, we consider the possibility that large galaxy surveys in the low-redshift Universe may be used to probe inhomogeneous reionization through its feedback on early galaxy formation. The many uncertainties remaining in our understanding of galaxy formation make it difficult to develop a rigourous formalism for this imprint, and motivate a simpler, hopefully more robust, approach. 

Besides the possibility of detecting reionization, its imprint in the galaxy power spectrum may act as a source of noise when probing cosmology.  Modern observations of the cosmic microwave background (CMB) \citep{deBernardis:2000,Halverson:2002,Mason:2003,Benoit:2003,Goldstein:2003,Spergel:2003} have greatly extended our knowledge of cosmological parameters.  One result has been the realisation that $\sim70\%$ of the Universe is composed of an unknown form of energy that generates the accelerated expansion seen in SN Ia observations \citep{riess1998,perlmutter1999}.  This is one of the most puzzling discoveries of our times, and it is hoped that future observations in the fields of SN Ia \citep{riess1998,riess2004,perlmutter1999}, weak lensing \citep{hoekstra2005}, and galaxy surveys \citep{se2003probe,blake2003} will constrain the time evolution of the dark energy giving clues as to its nature. For this reason, in this paper, we will focus on how reionization may affect estimates of dark-energy parameters.

Large galaxy surveys contribute information on the dark energy in two main ways.  First the form of the matter power spectrum, probed by galaxy surveys via the proxy of the galaxy power spectrum, depends on different parameter combinations than the CMB, breaking many of the parameter degeneracies \citep*{EHT1999}.  Secondly, the pre-recombination oscillation of the photon-baryon fluid leaves an imprint in the matter power spectrum, which may be used as a standard ruler to determine the angular diameter distance $D_A(z)$ as a function of redshift $z$ \citep{se2003probe,blake2003}.  These baryon oscillations have now been detected \citep{cole2005,eisenstein2005} by both 2dF and SDSS.  If the imprint in the galaxy power spectrum from patchy reionization can mimic or conceal any feature of the galaxy power spectrum from density fluctuations, then our ability to constrain dark energy using galaxy surveys will be degraded.

In this paper, we explore the possible consequences of this environmental dependence on the galaxy power spectrum.  The process of galaxy formation is still only poorly understood and so a detailed analysis of feedback is inappropriate.  Instead we choose to link galaxy formation to the neutral fraction by a simple {\em ansatz}, by which we hope to bring out the underlying behaviour, leaving the details for a later age.  In keeping with this ``simple is best" ideology, we choose to model the variation in neutral fraction using an analogue of the halo model \citep{cooray2002halo}.  With this approach we hope to phrase the problem in a general fashion, avoiding detailed assumptions about the reionization history.  To address these questions in a quantitative fashion, we employ the Fisher-matrix formalism (\citealt{jungman1996,jungman1996L}; \citealt*{TTH1997}).  This allows us to convert a theoretical model into predictions for the parameter constraints attainable by imagined experiments. 

The outline of this paper is as follows.  In \S \ref{sec:bubble} we detail the form of the ionization power spectrum and describe our simple {\em ansatz} relating it to galaxy formation.  Then, in \S \ref{sec:power}, we bring the two together describing the complete model galaxy power spectrum, including the effects of redshift distortions and the Alcock-Paczynski effect.  In \S \ref{sec:fisher} we outline the Fisher-matrix formalism.   Having set out our model, in \S \ref{sec:detect} we discuss the possibility of detecting reionization using galaxy surveys.  This is then expanded to consider the implications for dark-energy constraints in \S \ref{sec:dark}.  Finally, in \S \ref{sec:conclude} we summarise our conclusions.  

\section{Bubble model} 
\label{sec:bubble}

We wish to relate the overdensity of galaxies to the ionization fraction within a given region.  To do this, we make the simple {\em ansatz} that there is a component to the galaxy power spectrum which linearly traces the ionized fraction.  Hence, we may write the number density of galaxies $n(\mathbf{r})$ at position $\mathbf{r}$ as
\begin{equation}\label{num_density}
n(\mathbf{r})=\bar{n}[1+\delta_{\rm{gal}}(\mathbf{r})+\delta_{\rm{bub}}(\mathbf{r})],
\end{equation}
where $\bar{n}$ is the mean number density of galaxies, $\delta_{\rm{gal}}(\mathbf{r})=b\delta(\mathbf{r})$ assumes galaxies trace the underlying dark-matter fluctuations $\delta$ with bias $b$, and we calculate the fractional overdensity of galaxies due to an ionization field $x_i(\mathbf{r})$ by $\delta_{\rm{bub}}(\mathbf{r})=-\epsilon_b x_i(\mathbf{r})$, where $\epsilon_b$ parametrizes the strength of the effect.

Writing the number density in the form of equation \eqref{num_density} leads to a galaxy power spectrum
\begin{equation}
P(k)=\frac{1}{(1-\epsilon_b\bar{Q})^2}\left[P_{\rm{gal}}(k)+2P_{\rm{gal,bub}}(k)+P_{\rm{bub}}(k)\right],
\end{equation}
where $\bar{Q}$ is the filling fraction of the bubbles.  For simplicity, we choose to neglect the cross-correlation, which will be smaller or comparable in size to the other terms and represents an unnecessary refinement given the simplicity of our toy model.  Note the overall rescaling of the power spectrum because the mean galaxy density $\langle n\rangle=\bar{n}(1-\epsilon_b\bar{Q})$.  Typically, $\epsilon_b\bar{Q}\ll 1$ and we can neglect this correction and take
\begin{equation}
P(k)=P_{\rm{gal}}(k)+P_{\rm{bub}}(k).
\end{equation}

We now need to calculate the bubble power spectrum $P_{\rm{bub}}(k)$. In order to phrase the problem as broadly as possible, we eschew detailed assumptions about reionization in favour of a more general approach.  In this paper, we choose to associate regions of ionization with ``bubbles," in analogous fashion to the halo model's association of mass with haloes. Following the halo-model formalism \citep{cooray2002halo}, $P_{\rm{bub}}(k)$ is given by the sum of two terms, $P_{\rm{bub}}(k)=P^{1b}(k)+P^{2b}(k)$, which describe correlations within the same bubble and between two different bubbles respectively.  These terms are given by 
\begin{equation}
P^{1b}(k)=\epsilon_b^2\int dm\,n(m)\left(\frac{m}{\bar{\rho}}\right)^2|u(k|m)|^2,
\end{equation}
\begin{multline}
P^{2b}(k)=\epsilon_b^2\int dm_1\,n(m_1)\left(\frac{m_1}{\bar{\rho}}\right)u(k|m_1)\\
\times\int dm_2\,n(m_2)\left(\frac{m_2}{\bar{\rho}}\right)u(k|m_2)P_{bb}(k|m_1,m_2),
\end{multline}
where $n(m)$ is the comoving number density of bubbles of mass $m$, $P_{bb}(k|m_1,m_2)$ is the power spectrum of bubbles of mass $m_1$ and $m_2$, and $u(k|m)$ is the Fourier transform of the bubble ionization profile $u(\mathbf{r}|m)$.  With this notation, we may write the volume filling factor of the bubbles as $\bar{Q}=\int dm\,n(m)(m/\bar{\rho})$, and the bubble volume as $V_{\rm{bub}}=m/\bar{\rho}=4\pi r_{\rm{bub}}^3/3$, where $r_{\rm{bub}}$ is the comoving bubble radius.  Throughout this paper, we will assume a top-hat profile $u(\mathbf{r}|m)=\Theta(|\mathbf{r}|-r_{\rm{bub}}))/V_{\rm{bub}}$, for which $u(k|m)=3j_1(k r_{\rm{bub}})/(kr_{\rm{bub}})$, where $j_\ell(x)$ is a spherical Bessel function of order $\ell$.  

If we assume a delta-function size distribution and treat the power spectrum of the bubbles as tracing the dark-matter power spectrum $P_{bb}(k)\approx P_{\delta\delta}(k,z=z_{ri})$, where $z_{ri}$ is the redshift at which the imprint is formed, this reduces to
\begin{equation}
P^{1b}(k)=\epsilon_b^2\bar{Q}V_{\rm{bub}}|u(k|m)|^2,
\end{equation}
\begin{equation}
P^{2b}(k)=\epsilon_b^2\bar{Q}^2|u(k|m)|^2P_{\delta\delta}(k|m).
\end{equation}
In order to keep our model simple, we ignore evolution in the bubble-size distribution.  In reality, the relevant bubble sizes will be determined by the period when most baryons condense, an extended process that will average over the evolution of bubble growth.  We also ignore the effects of bubble overlap, which is expected to occur for large $\bar{Q}$ and undermines the halo-model approach.  Once bubbles begin to overlap, using isolated spheres to model the HII regions will not correctly represent the true size and shape of the ionized regions.  To a first approximation though, this effect will give an effective distribution of bubble sizes, and so should not affect our qualitative conclusions. We will take $\bar{Q}=0.5$ and $z_{ri}=6$ in what follows, and use $(r_{\rm{bub}},\epsilon_b)$ to parametrize the bubble power spectrum.  Note that the 2-bubble term is subdominant in the regime that we consider, making the details of $z_{ri}$, and any biasing of $P_{bb}(k)$ with respect to $P_{\delta\delta}(k)$, unimportant. 

\begin{figure}
\begin{center}
\resizebox{8cm}{!}{\includegraphics{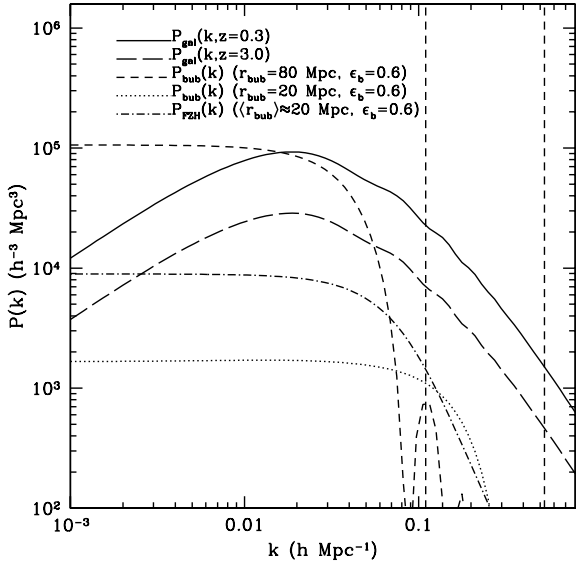}}\\%
\caption{
Comparison of the galaxy and bubble power spectra.  We plot $P_{\rm{gal}}(k)$ at two redshifts $z=0.3$ (solid curve) and $z=3.0$ (long dashed curve).  For each redshift, we plot the non-linear scale: $k_{\rm{max}}(z=0.3)=0.11 h\, \rm{Mpc^{-1}}$ and $k_{\rm{max}}(z=3.0)=0.53 h\, \rm{Mpc^{-1}}$ (dashed vertical lines, from left to right).  For comparison, we plot $P_{\rm{bub}}(k)$ for the parameters ($r_{\rm{bub}}=80\,\rm{Mpc},\epsilon_b=0.6$) (short dashed curve) and ($r_{\rm{bub}}=20\,\rm{Mpc},\epsilon_b=0.6$) (dotted curve).  Notice how the latter curve resembles constant white noise in the region $k<0.1h\, \rm{Mpc^{-1}}$.  The former curve displays a cutoff in power close to the galaxy-power-spectrum peak. Finally, we plot a bubble power spectrum $P_{\rm FZH}(k)$ (dot-dashed curve) that has been calculated using a bubble size distribution taken from \citet{fzh2004}, with $\langle r_{\rm bub}\rangle\approx20\,{\rm Mpc}$.
}
\label{fig:bubble_power}
\end{center}
\end{figure}

Figure \ref{fig:bubble_power} shows the form of the bubble power spectrum in this model.  Note that the power is fairly constant on small $k$ and cuts off sharply on linear scales smaller than the bubble radius.   As a simple example of including a smooth bubble size distribution, Figure \ref{fig:bubble_power} shows $P_{\rm bub}$ calculated using the bubble distribution of \citet{fzh2004}, assuming an ionizing efficiency $\zeta=40$ and $\bar{Q}=0.83$, which gives a volume averaged bubble size $\langle r_{\rm bub}\rangle\approx20\,{\rm Mpc}$.  The main effects of the distribution in bubble sizes are to smooth out the oscillations seen in the single size model and to decrease the rate at which power decreases on scales below the characteristic bubble size.   Having shown that the high-$k$ cutoff occurs even with a smooth distribution of bubble sizes, we will henceforth restrict ourselves to the simpler, single bubble size case.

The onset of non-linearity limits the scales that a galaxy survey is able to probe.  We choose to define this cutoff scale by requiring that the average fluctuation on a scale $R$ satisfies $\sigma(R)\le0.5$ for $R=\pi/(2k_{\rm{max}})$ \citep{se2003probe}.  This cutoff can lead to a degeneracy between the bubble spectrum and the constant shot-noise expected on large scales from non-Gaussian clustering of the galaxies \citep{seljak2000}.  We see in Figure \ref{fig:bubble_power} that on large scales the bubble power spectrum becomes constant.  If, for a given galaxy survey, $r_{\rm{bub}}$ is sufficiently small, then the curvature of the bubble spectrum will lie at $k>k_{\rm{max}}$, and the bubble spectrum will be indistinguishable from shot-noise.  Including galaxy surveys at higher $z$, where the non-linear scale is smaller, helps break this degeneracy.

We note that, for a random variable with zero mean, we would expect the power spectrum to vanish on large scales.  That this does not occur relates to a generic problem of the halo model 1-halo term, which is constant on large scales. In most applications this is masked by a dominant 2-halo term, which does decrease on large scales.  However, in our model the 2-bubble term is negligible making this issue obvious. Given that our model predicts a bubble power spectrum that looks like shot-noise on large scales, we must worry both about how the removal of shot-noise will affect our results and how to distinguish the effect of bubbles from shot-noise. As mentioned above, the existence of a cutoff in the bubble power spectrum distinguishes it from shot-noise (although we must observe this cutoff for this to work).  When we come to analyse the effect of bubbles on cosmological parameter estimation, we will include a term representing white shot-noise to account for this possible confusion. 

Having generated an imprint from patchy reionization, we must evolve it to lower redshift.  Our knowledge of how mergers recycle matter from many smaller haloes into fewer more massive haloes is not sufficient to handle this rigorously.  Instead, we will consider three cases that ought to bracket the truth.  We take $r_{\rm{bub}}$ to be a constant, fixing the shape of $P_{\rm{bub}}(k)$, and then consider how its amplitude varies with time.  We will consider three models for this time evolution
\[
P_{\rm{bub}}(k,z)=\left\{
\begin{array}{lc}
 P_{\rm{bub}}(k,z=z_{ri}), & (\rm{Model\,A}),  \\
 P_{\rm{bub}}(k,z=0) \left[\frac{G(z)}{G(z=0)}\right]^{2},&  (\rm{Model\,B}),  \\
 P_{\rm{bub}}(k,z=z_{ri}) \left[\frac{G(z)}{G(z=z_{ri})}\right]^{-2},& (\rm{Model\,C}).    
\end{array}
\right.
\]
In Model A, we assume that, once produced, the power spectrum $P_{\rm{bub}}(k)$ from bubbles  remains constant in time.  As the density fluctuations continue to grow this means that $P_{\rm{bub}}$ becomes less significant at later times.  In Model B, we allow $P_{\rm{bub}}(k)$ to grow as the square of the linear growth function $G(z)$.  Thus in Model B, $P_{\rm{bub}}(k)$ remains a constant fraction of the total galaxy power spectrum.  It seems unlikely that the bubble imprint would grow in this fashion, but we include this model in order to consider the case where the bubble imprint is equally important at all redshifts.  Note that in this model, we choose to normalise the bubble spectrum to the present day.  This provides a simple way of restricting $P_{\rm{bub}}(k)$ to amplitudes comparable to the density power spectrum. Finally, with Model C, we consider the case where $P_{\rm{bub}}(k)$ decreases with time.  This will provide an estimate of the worst-case scenario for detecting the bubbles. The time evolution of $P_{\rm{bub}}(k)$ is most important when we can compare surveys at different redshifts.  In the case of a single redshift survey, any growth can be absorbed into an effective $\epsilon_b$ for that survey.

What range of values may our two free parameters $\epsilon_b$ and $r_{\rm{bub}}$ reasonably take?  The characteristic size of the bubbles is the easiest question to address. \citet{fzh2004} present a model for bubbles forming around highly biased regions leading to typical sizes of $\sim5\,\rm{Mpc}$, when $\bar{Q}=0.5$ (see also \citealt*{fqh2006}).  In contrast, \citet{wyithe2004} use arguments based on light-travel times and cosmic variance to obtain bubble sizes of $\sim 60\,\rm{Mpc}$ at the end of reionization.  This latter value can be taken as an upper limit on reasonable bubble sizes, while the former gives a more reasonable estimate of what we might expect.  These values are in broad agreement with the results of computer simulation \citep{iliev2005,zahn2006}, which yield sizes $\sim 10\,\rm{Mpc}$.

The range of $\epsilon_b$ begs the question of how exactly to interpret this parameter.  We have assumed a linear relation between the ionization fraction of a region and its galaxy overdensity.  
We can readily see that $n(\mathbf{r})\ge0$, which implies a solid upper limit of $\epsilon_b\le1$.  An alternative approach is to consider the suppression of galaxy formation in haloes of low mass. Simulations at low redshift ($z<3$) \citep{thoul1996,kitayama2000} indicate significant suppression of galaxy formation in haloes with circular velocities $V_{\rm{circ}}\le 50 \kmis$.  At higher $z$, photoionization is less effective due to the decreased cooling time, decreased UV flux, increased self-shielding from the higher densities, and collapse beginning before any UV background can be generated  \citep{dijkstra2004}.  In this case, \citet{dijkstra2004} find that only haloes with $V_{\rm{circ}}\le 20 \kmis$ suffer significantly reduced condensation of baryons.  To estimate the mass fraction in galaxies affected by photoionization feedback, we take this latter value and apply it to the Press-Schecter distribution \citep{press1974} as a low-mass cutoff below which no galaxies form.  This gives an estimate of the decrement in galaxies due to photoionization feedback, 
\begin{equation}\label{epapprox}
\epsilon_b\approx\bar{\Delta}_g\equiv\left[\frac{F(M>M_{\rm{feedback}})-F(M>M_{\rm{cool}})}{F(M>M_{\rm{cool}})}\right],
\end{equation}
where $F(M)$ is the fraction of mass in haloes of mass greater than $M$, $M_{\rm{feedback}}$ is the mass corresponding to $V_{\rm{circ}}= 20 \kmis$, and $M_{\rm{cool}}$ is the mass corresponding to the virial temperature $T_{\rm{vir}}\approx10^4\,\rm{K}$ needed for effective cooling by atomic hydrogen.  Evaluating equation \eqref{epapprox} gives $\epsilon_b\approx 0.18$ at $z=10$ and $\epsilon_b\approx 0.10$ at $z=6$, which give an indication of sensible values.  Once a galaxy grows large enough, gravity will overcome feedback of this form and damp this effect.  Thus, these numbers represent an effective upper limit in the most plausible model.

\section{Galaxy power spectrum} 
\label{sec:power}
In constructing our galaxy power spectrum, we follow \citet{se2003probe}.  Incorporating the effects of bias, linear redshift distortions \citep{kaiser1987}, and linear growth, the galaxy power spectrum takes the form
\begin{equation}
P_{\rm{gal}}(k,\mu,z)=\left[\frac{G(z)}{G(z=0)}\right]^2b^2 (1+\beta \mu^2)^2 P_\delta(k,z=0),
\end{equation}
where $P_\delta(k,z=0)$ is the power spectrum of the dark matter at the present day, $b=\Omega_m(z)^{0.6}/\beta$ is the bias, and $\mu^2=k^2_{||}/k^2$ is the direction cosine between the Fourier-mode wavenumber and the line of sight.  We define the redshift-distortion parameter $\beta$ in terms of $\sigma_{8,g}$ and $\sigma_8$, the fluctuations in galaxies and dark matter, respectively, smoothed on scales of $8h^{-1}\,\rm{Mpc}$, by the relation,
\begin{equation}
\sigma_{8,g}=\sigma_8 b\sqrt{1+2\beta/3+\beta^2/5}.
\end{equation}
In order to calculate the linear growth factor $G(z)$, we integrate the perturbation equation,
\begin{equation}
\ddot{G}(t)+2H\dot{G}(t)-4\pi G\rho_{m}G(t)=0,
\end{equation}
with
\begin{equation}
\frac{H^2}{H_0^2}=\Omega_m(1+z)^{3}+(1-\Omega_m-\Omega_X)(1+z)^{2}+\Omega_X(z),
\end{equation}
and where the energy density in dark energy is given by
\begin{equation}
\Omega_X(z)=\Omega_X\exp\left(3\int_0^zdz'\,\frac{1+w(z')}{1+z'}\right).
\end{equation}
In the special case of a cosmological constant, the growth factor may be expressed as
\begin{equation}
G(z)=\frac{5}{2}\Omega_m\frac{H(z)}{H_0}\int_\infty^z \frac{1+z'}{[H(z')/H_0]^3}dz',
\end{equation}
but for a general dark-energy model where $w(z)\ne-1$, the full numerical integration is necessary \citep{wang1998,weinberg2003}.

Figure \ref{fig:bubble_power} shows $P_{\rm{gal}}(k,\mu,z)$ averaged over angle and evaluated at $z=0.3$ and $z=3$.  It displays a clear peak at $k\approx 0.02 h\, \rm{Mpc}^{-1}$, corresponding to the scale of matter-radiation equality, and visible baryon oscillations on smaller scales.  These features arise from the acoustic oscillation of the baryon-photon fluid during the period of tight coupling before recombination.  The sound speed, which governs the peak positions, is well measured from the CMB.  Consequently, the baryon oscillations may be used as a standard ruler, allowing a direct measurement of the angular-diameter distance.  These features have now been detected in galaxy surveys \citep{cole2005,eisenstein2005}, and their use in probing the dark energy is well known \citep{se2003probe,blake2003}.

When converting the observed redshift and angular position of galaxies into linear space, we must assume a particular cosmology.  If this reference cosmology is different from the true cosmology, then we will introduce distortions into the inferred distribution of galaxies.  This is the 
Alcock-Paczynski (AP) effect \citep{alcock1979} and is essentially a cosmological redshift distortion.
We may express the power spectrum inferred by our observations in terms of the true power spectrum $P^{\rm{tr}}(k^{\rm{tr}},\mu^{\rm{tr}})$ by
\begin{equation}
P_{\rm{obs}}(k,\mu)=\frac{D_A^2(z)H^{\rm{\rm{tr}}}(z)}{D_A^{\rm{\rm{tr}}\,2}(z)H(z)}P^{\rm{tr}}(k^{\rm{\rm{tr}}},\mu^{\rm{\rm{tr}}}),
\end{equation}
where $H$ and $D_A$ are calculated using the reference cosmology, and $H^{\rm{tr}}$ and $D_A^{\rm{tr}}$ with the true cosmology.  We write the components of a Fourier wavevector parallel and perpendicular to the line of sight as $k_{||}^{\rm{tr}}=(H^{\rm{tr}}/H) k_{||}$ and $k_{\perp}^{\rm{tr}}=(D_A/D_A^{\rm{tr}}) k_\perp$.  The information contained in the AP effect can be useful in probing the evolution of the dark energy, so we include it in this analysis.

The final observed galaxy power spectrum incorporates all of the effects that we have discussed before and takes the form
\begin{multline}\label{power_final}
P_{\rm{obs}}(k,\mu)=\frac{D_A^2(z)H^{\rm{\rm{tr}}}(z)}{D_A^{\rm{\rm{tr}}\,2}(z)H(z)}\left[P_{\rm{gal}}(k^{\rm{\rm{tr}}},\mu^{\rm{\rm{tr}}})+P_{\rm{bub}}(k^{\rm{\rm{tr}}})\right]\\+P_{\rm{shot}},
\end{multline}
where $P_{\rm{shot}}$ is residual shot-noise from non-Gaussian clustering of galaxies \citep{seljak2000}, which we treat as a constant white-noise term. 
\section{Fisher matrix} 
\label{sec:fisher}

To quantitatively constrain the effect of bubbles on the galaxy power spectrum, we turn to the Fisher matrix.  This formalism allows us to estimate the uncertainties on a set of model parameters $\Theta=(\theta_1, \theta_2,...,\theta_N)$ given some data set.  We define the Fisher matrix \citep{TTH1997}
\begin{equation}
F_{ij}\equiv -\left\langle \frac{\partial^2 \log L}{\partial\theta_i\partial\theta_j}\right\rangle\bigg|_{\Theta_0},
\end{equation}
where $L$ is the likelihood function describing the probability distribution of the parameters and $\Theta_0$ is the place in parameter space where the Fisher matrix is evaluated, typically the point of maximum likelihood.  Given the Fisher matrix, the Cramer-Rao inequality states that the minimum uncertainty on a parameter $\theta_i$ is given by $\Delta\theta_i\ge (F^{-1})_{ii}^{1/2}$. 
This estimate of the uncertainty will be reliable provided that $\Theta_0$ is near to the true values of the parameters.

To evaluate $F_{ij}$, we need to specify a model, which determines the dependence of the likelihood function on $\Theta$, and a point in parameter space where we wish to determine parameter uncertainties.   In the case that the model parameters are Gaussian distributed, the Fisher matrix takes the form 
  \begin{equation}\label{fisher_general}
F_{\alpha\beta}=\frac{1}{2}\rm{\rm{tr}}(C^{-1}C,_\alpha C^{-1}C,_\beta)+\frac{\partial\mu}{\partial\theta_\alpha}C^{-1}\frac{\partial\mu}{\partial\theta_\beta},
\end{equation}
where $C$ is the covariance matrix for the data, and $\mu$ is the data's mean.  This will be a good approximation in the case of both CMB observations and galaxy survey.  Note that, for our purposes, we will need to combine information from both the CMB and galaxy surveys.  When used together these data sets break many degeneracies that are present when they are used alone.  Let us consider the Fisher matrix from each of these in turn.

A CMB experiment may be characterised by a beam size $\theta_{\rm{beam}}$ and sensitivities to temperature $\sigma_T$ and polarization $\sigma_P$.  Given these quantities, the Fisher matrix is given by (\citealt{jungman1996,jungman1996L}; \citealt*{kks,zaldarriaga1997})
 \begin{equation}\label{fisher_cmb}
F^{CMB}_{\alpha\beta}=\sum_\ell\sum_{X,Y} \frac{\partial C_\ell^X}{\partial\theta_\alpha} (\rm{Cov}_\ell)^{-1}_{XY}\frac{\partial C_\ell^Y}{\partial\theta_\alpha},
\end{equation}
where $C_\ell^X$ is the power in the $\ell$th multipole for $X=T,E,B,$ and $C$,  the temperature, E-mode polarization, B-mode polarization, and TE cross-correlation respectively.  The elements of the covariance matrix $\rm{Cov}_\ell$ between the various power spectra are \citep{kks,zaldarriaga1997}
\begin{eqnarray}
(\rm{Cov}_\ell)_{TT}&=&\frac{2}{(2\ell+1)f_{\rm{sky}}}(C_{T\ell}+w_T^{-1}B_\ell^{-2})^2,\nonumber\\
(\rm{Cov}_\ell)_{EE}&=&\frac{2}{(2\ell+1)f_{\rm{sky}}}(C_{E\ell}+w_P^{-1}B_\ell^{-2})^2,\nonumber\\
(\rm{Cov}_\ell)_{BB}&=&\frac{2}{(2\ell+1)f_{\rm{sky}}}(C_{B\ell}+w_P^{-1}B_\ell^{-2})^2,\nonumber\\
(\rm{Cov}_\ell)_{CC}&=&\frac{1}{(2\ell+1)f_{\rm{sky}}}[C^2_{C\ell}+(C_{T\ell}+w_T^{-1}B_\ell^{-2})\nonumber\\
&&\times(C_{T\ell}+w_T^{-1}B_\ell^{-2})],\nonumber\\
(\rm{Cov}_\ell)_{TE}&=&\frac{2}{(2\ell+1)f_{\rm{sky}}}C_{C\ell}^2,\nonumber\\
(\rm{Cov}_\ell)_{TC}&=&\frac{2}{(2\ell+1)f_{\rm{sky}}}C_{C\ell}(C_{T\ell}+w_T^{-1}B_\ell^{-2}),\nonumber\\
(\rm{Cov}_\ell)_{EC}&=&\frac{2}{(2\ell+1)f_{\rm{sky}}}C_{C\ell}(C_{E\ell}+w_P^{-1}B_\ell^{-2}),\nonumber\\
(\rm{Cov}_\ell)_{TB}&=&(\rm{Cov}_\ell)_{EB}=(\rm{Cov}_\ell)_{CB}=0.
\end{eqnarray} 
Here $B_\ell^2$ is the beam window function, assumed Gaussian with $B_\ell^2=\exp[-\ell(\ell+1)\theta_{\rm{beam}}^2/8\ln 2]$, where $\theta_{\rm{beam}}$ is the full-width, half-maximum (FWHM) of the beam in radians.  Also, $w_T$ and $w_P$ are the inverse square of the detector noise for temperature and polarization, respectively.  For multiple frequency channels we replace $w_T B^2_\ell$ with the sum of this quantity for each of the channels.  

Moving now to galaxy surveys, we may write the appropriate Fisher matrix as
\citep{tegmark1997}
\begin{equation}\label{fisher_gal}
F^{GAL}_{\alpha\beta}=\int_0^{k_{\rm{\rm{max}}}}\frac{\partial\ln P(\mathbf{k})}{\partial\theta_\alpha}\frac{\partial\ln P(\mathbf{k})}{\partial\theta_\beta}V_{\rm{eff}}(\mathbf{k}) \frac{d^3\mathbf{k}}{2(2\pi)^3},
\end{equation}
where the derivatives are evaluated using the cosmological parameters of the fiducial model and $V_{\rm{eff}}$ is the effective volume of the survey, given by
\begin{eqnarray}
V_{\rm{eff}}(k,\mu)&=&\int d^3\mathbf{r}\left[\frac{n(\mathbf{r})P(k,\mu)}{n(\mathbf{r})P(k,\mu)+1}\right]^2\nonumber\\
&=&\left[\frac{\bar{n}P(k,\mu)}{\bar{n}P(k,\mu)+1}\right]^2V_{\rm{survey}}.
\end{eqnarray}
Here the galaxy survey is parametrized by the survey volume $V_{\rm{survey}}$ and the galaxy density $n(\mathbf{r})$, which in the last equality we assume to be uniform $\bar{n}$.  In addition, we must specify $k_{\rm{max}}$, a cutoff on small scales to avoid the effects of non-linearity.  We choose to define this cutoff scale by the criterion $\sigma(R)\le0.5$ for $R=\pi/(2k_{\rm{max}})$ \citep{se2003probe}. 

To apply the above framework, we need a theory relating the observables $C^X_l$ and $P(k,\mu)$ to the parameters.  For the CMB this is standard, while in the case of the galaxy surveys we use equation \eqref{power_final}, which arose from our discussion in \S \ref{sec:bubble} and \S\ref{sec:power}.  Using these models, we calculate the Fisher matrices for individual galaxy surveys and our CMB experiment, and then combine them
\begin{equation}
F_{\alpha\beta}^{TOT}=F_{\alpha\beta}^{CMB}+\sum_i F_{\alpha\beta}^{GAL,i},
\end{equation}
where $i$ labels the different galaxy surveys. This total Fisher matrix is then inverted to get parameter error predictions.

For this calculation, we need the specifications of our experiments.  These are given in Tables \ref{tab:CMBdata} and \ref{tab:GALdata}.  We consider two galaxy surveys.  The first uses parameters corresponding to the SDSS LRG survey, which is currently underway.  The second is a hypothetical survey at $z=3$, based upon a survey of Lyman break galaxies \citep{se2003probe}.
\begin{table}
\caption{Specification for CMB experiments}
\begin{center}
\begin{tabular}{|c|c|c|c|c|}
\hline
Experiment & Frequency& $\theta_{\rm{beam}}$ & $\sigma_T$ & $\sigma_P$ \\
\hline
WMAP & 40 & 28.2 &17.2 & 24.4 \\
	& 60 & 21.0 & 30.0 & 42.6 \\
	& 90 & 12.6 & 49.9 & 70.7 \\
\hline
Planck & 143 & 8.0 & 5.2 & 10.8 \\
	& 217 & 5.5 & 11.7 & 24.3 \\
\hline
\end{tabular}
\end{center}
Notes: Frequencies are in GHz.  Beam size $\theta_{\rm{beam}}$ is the FWHM in arcsec.  Sensitivities $\sigma_T$ and $\sigma_P$ are in $\mu \rm{K}$ per FWHM beam, $w=(\theta_{\rm{beam}}\sigma)^{-2}$. Taken from \citet{EHT1999}.
\label{tab:CMBdata}
\end{table}%

\begin{table}
\caption{Specification for galaxy surveys}
\begin{center}
\begin{tabular}{|c|c|c|c|c|c|}
\hline
 Survey& $z$ & $V_{\rm{survey}}$ & $\bar{n}$& $k_{\rm{\rm{max}}}$ & $\sigma_{8,g}$\\
 	&	&$(h^{-3} \rm{Gpc)^3}$&$(h^3\, \rm{Mpc}^{-3})$&$(h\, \rm{Mpc}^{-1})$&\\
\hline
SDSS & 0.3 & 1.0 &$10^{-4}$ & 0.11 &1.8\\
\hline
S2 & 3.0 & 0.50 & $10^{-3}$ & 0.53 &1.0\\
\hline
\end{tabular}
\end{center}
Notes: Taken from \citet{se2003probe}.
\label{tab:GALdata}
\end{table}%

Finally, we must decide upon our choice of parameter space.  We specify our cosmology using
7 parameters describing the total matter fraction $\Omega_m h^2$, $\Omega_m$, the baryon fraction $\Omega_b h^2$, the inflationary amplitude $A^2_S$, scalar spectral index $n_s$, optical depth to last scattering $\tau$, and the tensor-scalar ratio $T/S$; each galaxy survey is described by five parameters ($\log H,\log D_A,\log G,\log \beta,P_{shot}$); to these, we add two parameters ($\epsilon_b$, $r_{\rm{bub}}$) to describe our bubble model (we choose to set $\bar{Q}=0.5$ and $z_{ri}=6$).  In choosing these parameters, we are following \citet{se2003probe}.  We treat all of the above parameters as being independent and then extract information about the dark energy from our uncertainties on $(\log H,\log D_A)$ from each survey.  We choose to parametrize the dark energy using three parameters ($\Omega_X,w_0$,$w_1$), taking the dark-energy equation-of-state parameter to be $w(z)=w_0+w_1z$.  In deciding on our fiducial values, we follow the results of WMAP \citep{Spergel:2003} for the cosmological parameters. These are broadly consistent with the updated results of \citet{Spergel:2006}, except for the decreased $\Omega_m$ and $\tau$.  Evaluating the Fisher matrix at these different best-fitting parameters modifies our constraints only slightly. The bubble parameters are highly uncertain, and so we choose to explore a large parameter space.

\section{Possiblity of detecting bubbles} 
\label{sec:detect}

Now that we have established a theoretical framework, we wish to determine whether the imprint can be detected using the specified galaxy surveys.  Our null hypothesis is that there is no imprint, and we assume that a detection requires that we can distinguish both $\epsilon_b$ and $r_{\rm{bub}}$ from zero at approximately the $2\sigma$ level; i.e., we require both $\epsilon_b>2\sigma_{\epsilon_b}$ and $r_{\rm{bub}}>2\sigma_{r_{\rm{bub}}}$.  Throughout, we assume the inclusion of CMB information at the level of Planck.  Less precise CMB data will relax constraints on cosmological parameters, causing parameter degeneracies to decrease the sensitivity of the galaxy survey to the bubble imprint.
\begin{figure}
\begin{center}
\resizebox{8cm}{!}{\includegraphics{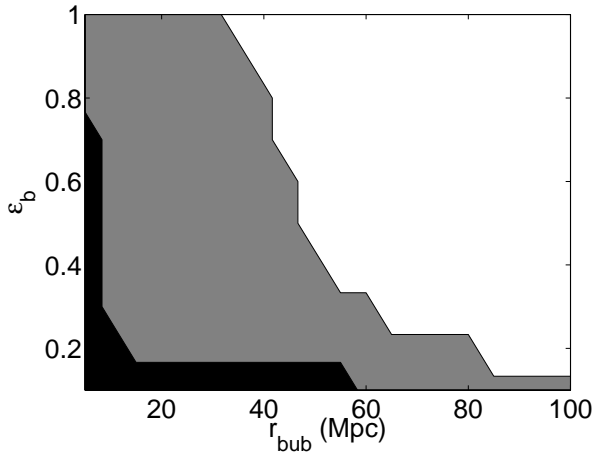}}\\%
\caption{Model A: Contour map of detection $\epsilon_b>2\sigma_{\epsilon_b}$ and $r_{\rm{bub}}>2\sigma_{r_{\rm{bub}}}$ in the bubble parameter plane.  The white region is detectable by SDSS alone, the grey region is detectable by SDSS+S2, and the black region is undetectable to all surveys.  Planck CMB data is assumed in all calculations.}
\label{fig:detectmap}
\end{center}
\end{figure}

\begin{figure}
\begin{center}
\resizebox{8cm}{!}{\includegraphics{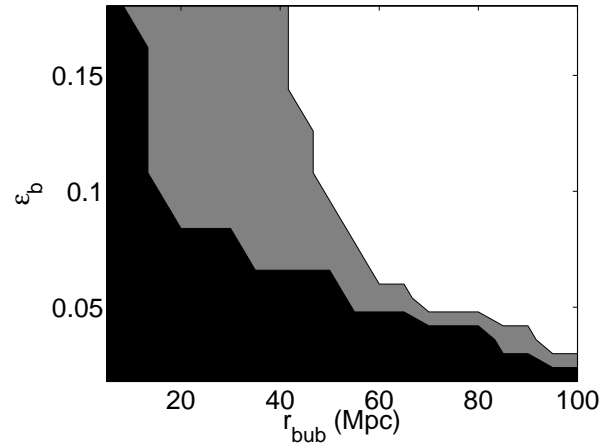}}\\%
\caption{Model B: Contour map of detection. As for Figure \ref{fig:detectmap}. For comparison with other figures, note that $G(z=6)/G(z=0)=0.18$, so that $[G(z=0)/G(z=6)]\epsilon_b$ lies in the range [0,1].}
\label{fig:time_detectmap}
\end{center}
\end{figure}

\begin{figure}
\begin{center}
\resizebox{8cm}{!}{\includegraphics{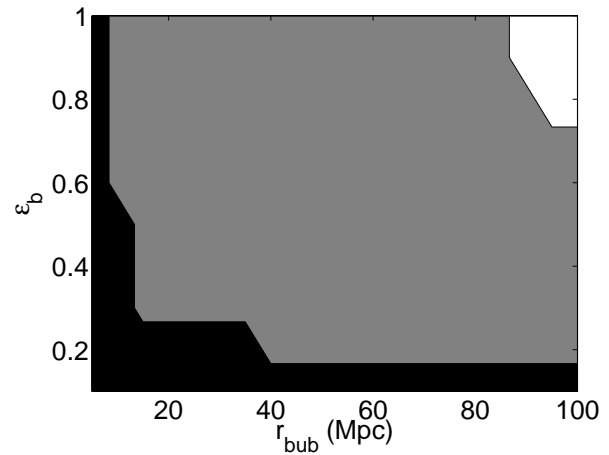}}\\%
\caption{Model C: Contour map of detection. As for Figure \ref{fig:detectmap}. Note the greatly decreased ability of SDSS alone to detect bubbles when compared with Figure \ref{fig:detectmap}.}
\label{fig:timeshrink_detectmap}
\end{center}
\end{figure}
Figures \ref{fig:detectmap}, \ref{fig:time_detectmap}, and \ref{fig:timeshrink_detectmap} show contour plots for models A, B, and C, denoting regions of parameter space where our surveys are able to make a detection. We shade the region of the ($r_{\rm{bub}},\epsilon_b$) plane where a detection can be made by SDSS alone (white), SDSS and S2 combined (grey), and where no detection can be made (black).  For models A and C, we consider the region of parameter space with $r_{\rm{bub}}\in[5\,\rm{Mpc},100\,\rm{Mpc}]$ and $\epsilon_b\in[0.1,1]$.  For model B, we choose a slightly different normalisation so that $[G(z=0)/G(z=6)]\epsilon_b\in[0.1,1]$.  This makes the range of amplitude of $P_{\rm{bub}}$ at $z=0$ identical in the region covered in Figures \ref{fig:detectmap} and \ref{fig:time_detectmap}.  Note that $G(z=6)/G(z=0)=0.18$ in our fiducial cosmology.

First, compare Figures \ref{fig:detectmap} and \ref{fig:timeshrink_detectmap}.  In model A, we see that SDSS alone is able to detect bubbles over a wide range of $\epsilon_b$, provided that the bubbles are large ($r_{\rm{bub}}>40\,\rm{Mpc}$).  In contrast, when we allow the bubble amplitude to decrease with time, as in Model C, we see that SDSS alone is almost unable to constrain either bubble parameter.  In both cases, addition of the S2 survey greatly improves the situation, allowing a wider range of parameter space to be probed.  However, even with S2, the theoretically preferred region with $r_{\rm{bub}}<10\,\rm{Mpc}$ and $\epsilon_b<0.2$, towards the bottom left hand corner, remains unconstrained.  The prospects for detection are clearly enhanced by including galaxy surveys at higher redshift.  

Figure \ref{fig:time_detectmap}, for Model B, shows that growth improves the prospects for probing smaller values of $\epsilon_b$, but makes little difference to our ability to constrain the bubble size.  As in model A, SDSS alone can only probe bubbles with $r_{\rm{bub}}>40\,\rm{Mpc}$, and S2 is required to probe smaller scales.  Note that when we normalize to the present day, the inclusion of growth in model B reduces the amplitude of the bubble imprint seen by the S2 survey by a factor of $[G(z=3)/G(z=0)]^2\approx0.1$ over that in model A.  This is responsible for the increased region that is undetectable to SDSS+S2 in Figure \ref{fig:time_detectmap}.  The amplitude of $P_{\rm{bub}}$ at $z=0.3$ is very similar 
in these two models, resulting in the nearly identical contours for SDSS only (when rescaled to account for the different normalisation) in Figures \ref{fig:detectmap} and \ref{fig:time_detectmap}. The striking differences between the three models indicates the importance of the time evolution of the bubble imprint.

It is worth pausing for a moment to consider where our leverage on the bubble power spectrum originates.  When we combine the two surveys, the bulk of the improvement is coming from the S2 survey alone.  This is unsurprising, as the growth of the density fluctuations means the bubble imprint is a more significant contribution to the galaxy power spectrum at early times.
Further, if we consider Figure \ref{fig:bubble_power}, we see that for very small bubble sizes, the bubble spectrum begins to resemble white noise over the region probed by the galaxy surveys.  This would further complicate detecting the bubble imprint as it could then be confused with residual Poisson shot-noise in the galaxy counts.  This problem is greatest at low $z$ where the non-linear scale is larger.  Both of these motivate performing this test in galaxy surveys at increasing redshift, ideally at the redshift of reionization, where an H$\alpha$ survey may be possible.

We conclude that galaxy surveys should be sensitive to the imprint in the galaxy power spectrum left over after reionization.  However, detecting this imprint will be difficult unless the characteristic size of HII regions is large ($r_{\rm{bub}}>10\,\rm{Mpc}$) and the effects of feedback significant ($\epsilon_b>0.1$).  This should be sufficient to constrain the more extreme models for reionization, but is unlikely to impact more reasonable scenarios. There is significant uncertainty in this prediction stemming from the difficulty in predicting the evolution of the imprint to more recent times.  

Currently, the best hope for measuring the size of HII regions during the early stages of reionization lies with upcoming 21cm observations (e.g., LOFAR\footnote{See http://www.lofar.org/.}, MWA\footnote{See http://web.haystack.mit.edu/arrays/MWA/.}, or PAST\footnote{See \citet*{pen2005past}}).  Direct imaging of the HII regions is unlikely with the first generation of detectors, but the prospects for statistical detection at $z\le10$ are good (\citealt*{zfh2004freq,bowman2006}; \citealt{mcquinn2005}).  At higher redshifts, $z>10$, corresponding to lower frequencies, sky noise increases dramatically making observations more difficult.  An imprint directly upon the galaxy power spectrum avoids these technical issues, making possible a complementary measurement.  In the event of very early reionization, detection of the imprint discussed in this paper might provide weak constraints on reionization before 21cm experiments reach the desired sensitivity. 

\section{Implications for dark-energy constraints} 
\label{sec:dark}

Current constraints on dark-energy parameters arise from the combination of high-precision CMB data with information from galaxy surveys.  The combination of high-$z$ ($z>3$) information, long before dark energy becomes dynamically important, with low-$z$ ($z<3$) information, deep within the dark-energy-dominated regime, serves to break many of the degeneracies that either data set possesses when used alone.  Adding in more galaxy surveys at different redshifts further constrains the evolution of the dark energy, allowing constraints on both $\Omega_X$ and its equation-of-state parameter $w(z)$.  In the previous section, we considered the bubble imprint as a useful signal; in this section we consider it as a potential source of noise for galaxy surveys.  If the bubble power spectrum is able to mimic the effects of dark energy, then it will degrade our ability to constrain dark-energy parameters.  Throughout this section, we will consider a dark-energy model with $w_0=-1$ and $w_1=0$.  Our numerical results depend upon this choice of model, but the overall picture remains the same when $w_0$ and $w_1$ take other values.

Galaxy surveys provide direct constraints on the dark energy through both the baryon oscillations and from the Alcock-Paczynski effect.  They also provide indirect constraints in combination with CMB data, as they probe $\Omega_m$ independently of $\Omega_mh^2$, the parameter directly probed by the CMB.  This allows the CMB indication of flatness $\Omega_k\approx0$ to constrain $\Omega_{X}$.  The bubble imprint must interfere with one of these measurements to be a source of confusion.

Measurement of the baryon oscillations allow a determination of the angular-diameter distance.  Their distinctive oscillatory structure is very different from the smooth structure that we expect from any plausible bubble imprint and so we do not expect there to be any confusion between the two.  The inferred peak position, amplitude, and overall shape of the galaxy power spectrum, on the other hand, could  be affected by the smooth form of the bubble imprint, making these the most likely points of confusion.  Thus, we expect parameters such as $\Omega_m$ and $n_s$ to be sensitive to the bubble imprint.  This simple picture is modified by inclusion of CMB data, which places tight constraints on many of these parameters making the effect of the bubble imprint more subtle.

\begin{figure}
\begin{center}
\setlength{\unitlength}{1cm}
\begin{picture}(1,8)
\put(0.3,5.5){ $\epsilon_b$}
\put(0.3,4.8){ $r_{\rm{bub}}$}
\put(0.3,4.0){ $A_S^2$}
\put(0.3,3.2){ $n_s$}
\put(0.3,2.5){ $w_1$}
\put(0.3,1.8){ $w_0$}
\put(0.3,1.0){ $\Omega_X$}
\put(0,0.3){ $\Omega_m h^2$}
\end{picture}
\resizebox{6cm}{!}{\includegraphics{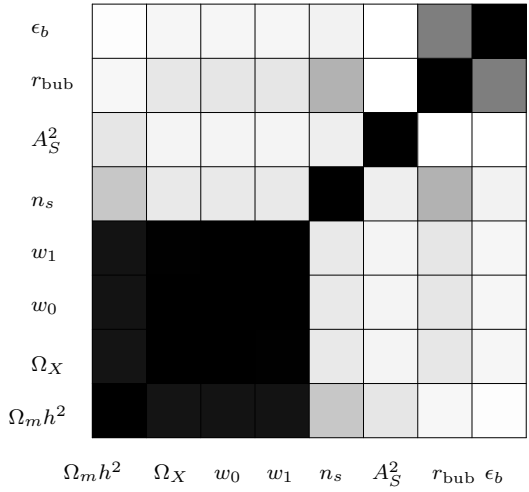}}\\%
\setlength{\unitlength}{1cm}
\begin{picture}(8,1)
\put(6.8,0.6){ $\epsilon_b$}
\put(6.1,0.6){ $r_{\rm{bub}}$}
\put(5.3,0.6){ $A_S^2$}
\put(4.6,0.6){ $n_s$}
\put(3.9,0.6){ $w_1$}
\put(3.2,0.6){ $w_0$}
\put(2.4,0.6){ $\Omega_X$}
\put(1.2,0.6){ $\Omega_m h^2$}
\end{picture}
\caption{Illustration of the reduced covariance matrix. ($\Omega_mh^2$,$\Omega_X$,$w_0$,$w_1$,$n_s$,$A_s^2$,$r_{\rm{bub}}$,$\epsilon_b$).  The model uses $r_{\rm{bub}}=10\,\rm{Mpc}$ and $\epsilon_b=0.1$.  Black indicates strong correlation and white indicates little correlation.}
\label{fig:fisherplot}
\end{center}
\end{figure}

The correlation between the different parameters is indicated in Figure \ref{fig:fisherplot}, for a model with $r_{\rm{bub}}=10\,\rm{Mpc}$ and $\epsilon_b=0.1$.  Note that there is a weak correlation between $r_{\rm{bub}}$ and the dark-energy parameters.  A slightly larger covariance is seen between $r_{\rm{bub}}$ and $n_s$.  At larger values of ($r_{\rm{bub}},\epsilon_b$), the picture remains unchanged except for a breaking of the degeneracy between $r_{\rm{bub}}$ and $\epsilon_b$ as the cutoff in $P_{\rm{bub}}(k)$ on small scales becomes more pronounced.

Before detailing the effect the bubble imprint has on statistical errors, let us consider the possibility of systematic biasing of our best-fitting values, if an existing bubble imprint was ignored in the analysis of data.  This will be relevant only in the case that the bubble imprint is not easily detectable, as an obvious imprint would certainly be included in the data analysis.  For the case where the bubbles are not detected -- i.e., $\epsilon_b<2\sigma_{\epsilon_b}$ and $r_{\rm{bub}}<2\sigma_{r_{\rm{bub}}}$ -- we have estimated this systematic offset between the inferred and true parameters, using the Fisher matrix to approximate the full likelihood surface.  We find that the offset is significantly smaller than the parameter uncertainty, typically being of order $\sim0.1\%$.   From this, we conclude that failing to include the imprint should not systematically affect parameter estimates in the near future.  When galaxy surveys begin to probe cosmological parameters below the percent level this effect will need to be included. We now turn to the effect of the imprint on parameter constraints.

\begin{figure}
\begin{center}
\resizebox{8cm}{!}{\includegraphics{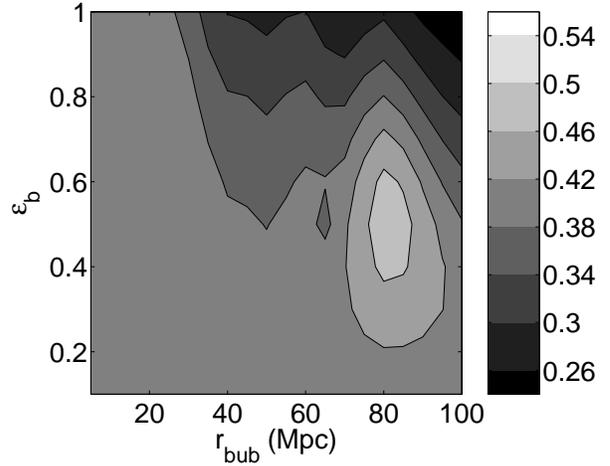}}\\%
\caption{Model A: Contour map of errors in $w_0$ in the bubble parameter plane.  We plot contours spanning the range $\sigma_{w_0}=0.22$--$0.58$ in intervals of $0.04$. The fiducial model takes $w_0=-1$ and $w_1=0$.  When no bubbles are present, we find $\sigma_{w_0}=0.39$. }
\label{fig:werrormap}
\end{center}
\end{figure}

\begin{figure}
\begin{center}
\resizebox{8cm}{!}{\includegraphics{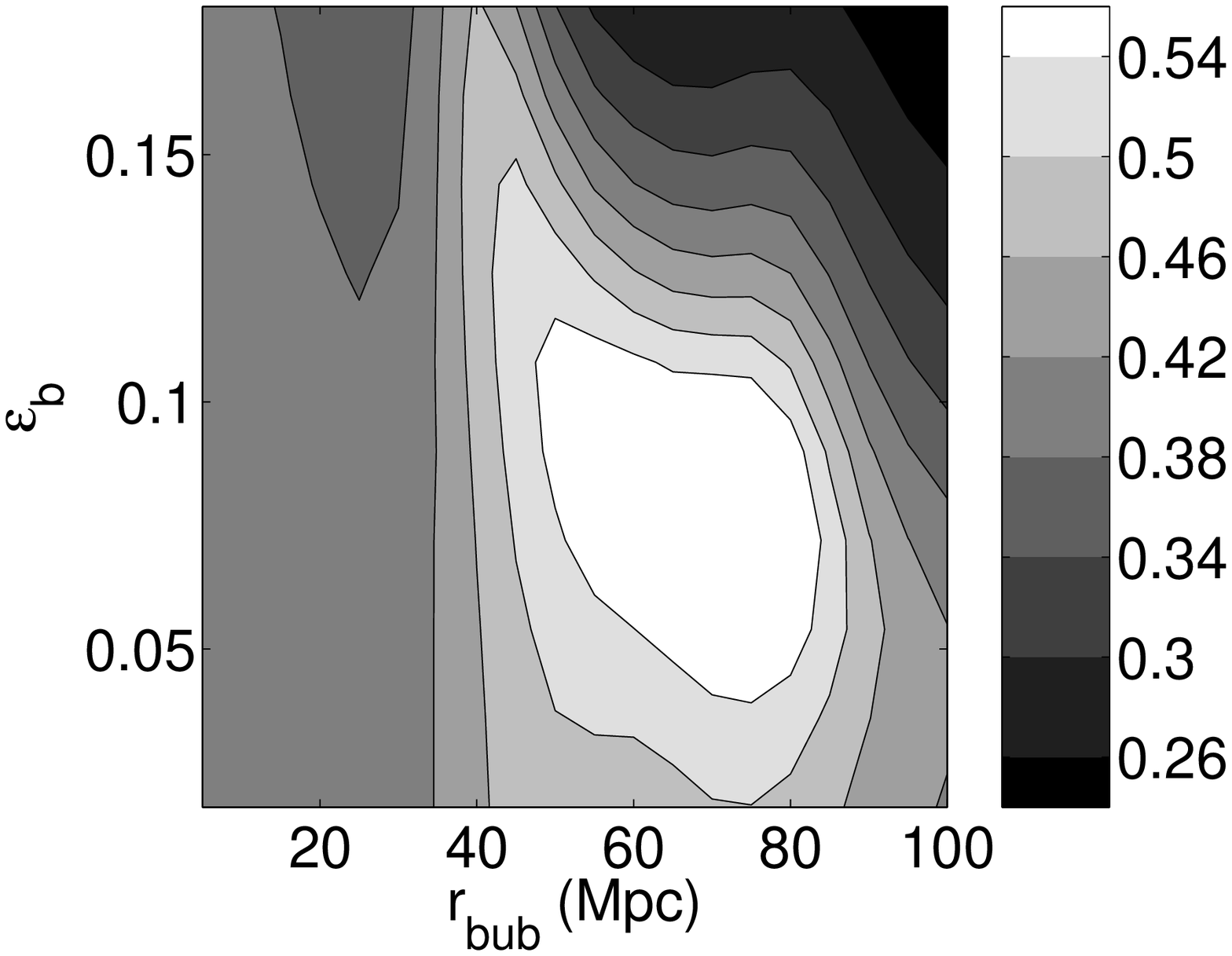}}\\%
\caption{Model B: Contour map of errors in $w_0$ in the bubble parameter plane. As for Figure \ref{fig:werrormap}.}
\label{fig:time_werrormap}
\end{center}
\end{figure}

\begin{figure}
\begin{center}
\resizebox{8cm}{!}{\includegraphics{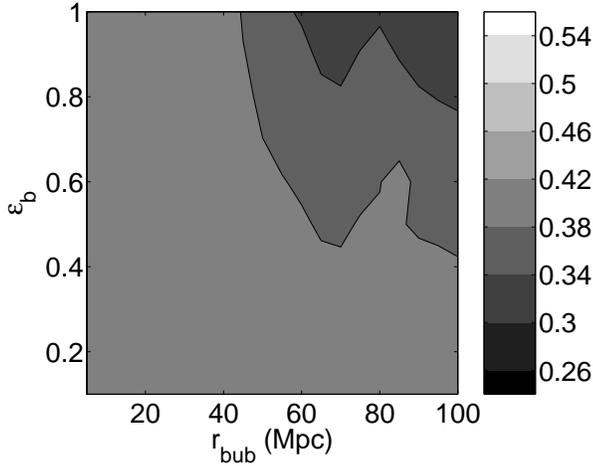}}\\%
\caption{Model C: Contour map of errors in $w_0$ in the bubble parameter plane. As for Figure \ref{fig:werrormap}.}
\label{fig:timeshrink_werrormap}
\end{center}
\end{figure}

Figures \ref{fig:werrormap}, \ref{fig:time_werrormap}, and  \ref{fig:timeshrink_werrormap} indicate error contours for $w_0$ over the ($r_{\rm{bub}},\epsilon_b$) plane.  The same shading scheme is used in all three figures to allow easy comparison.  First consider Figure \ref{fig:werrormap}. We see that the uncertainty on $w_0$ is maximal for bubble parameters $r_{\rm{bub}}\approx80\,\rm{Mpc}$ and $\epsilon_b\approx0.5$.  The form of $P_{\rm{bub}}(k)$ is plotted in Figure \ref{fig:bubble_power} where we see that it cuts off close to the maximum of the density power spectrum.  This is consistent with our above statements.  We find a maximum uncertainty of $\sigma_{w_0}=0.48$ in contrast with the uncertainty $\sigma_{w_0}=0.39$ in the absence of bubbles.  This indicates that bubbles can be an important source of noise in attempts to constrain dark energy.  However, the large values of ($r_{\rm{bub}},\epsilon_b$) required for this effect seem theoretically unlikely and from the discussion in \S\ref{sec:detect} would allow direct detection of the bubbles.  For more reasonable choices of bubble parameters ($r_{\rm{bub}}<10\,\rm{Mpc}$,$\epsilon_b<0.1$), the uncertainty on $w_0$ reduces to $\sigma_{w_0}=0.39$.  Thus, the effect of the bubble imprint is likely to be somewhat important in future attempts to constrain dark energy.  

Now consider Figures \ref{fig:time_werrormap} and  \ref{fig:timeshrink_werrormap}.  The increased uncertainty in $w_0$ is more pronounced in Model B, where the uncertainty rises as high as $\sigma_{w_0}=0.62$. Even here, for small values of $(r_{\rm{bub}},\epsilon_b$), the bubble imprint becomes unimportant and we recover $\sigma_{w_0}=0.39$, the no-bubble uncertainty.  Again this maximal uncertainty occurs close to $r_{\rm{bub}}\approx80\,\rm{Mpc}$ and $\epsilon_b\approx0.5$.  The increased value of $\sigma_{w_0}$ is a consequence of the bubble imprint growing at the same rate as the density fluctuations.  Consequently, the overall shape of the total galaxy power spectrum remains constant in time.  Thus, combining information at two redshifts provides much less leverage on separating out the two components, leading to larger parameter uncertainties.  In model C, the damping of the imprint means that it has much less effect on the dark-energy parameters.

Finally, we note that Figures \ref{fig:werrormap}, \ref{fig:time_werrormap}, and \ref{fig:timeshrink_werrormap} display a region of decreased uncertainty in $w_0$ in the top right hand corner, when the bubbles are large and feedback strong.  This is an interesting example of the AP effect.  In this region, the bubble power spectrum dominates over the density contribution and the overall shape of the galaxy power spectrum displays a well defined, sharp cutoff.  Distortion of this scale by the AP effect places good constraints on the dark energy. Galaxy surveys already show that the galaxy power spectrum closely traces the underlying density field, so this region is ruled out.

Having considered dark-energy parameters, it would seem natural to also consider inflationary parameters; e.g., the tilt $n_s$ and amplitude $A_S^2$.  For the surveys that we have analysed, inclusion of the bubble power spectrum makes little difference to the uncertainty on these parameters.  Essentially, all of the information needed for constraining these quantities is contained within the CMB.  In the absence of information on the optical depth $\tau$, or if there are significant tensor modes, galaxy-survey information becomes important in breaking degeneracies.  This is not true in the cases that we consider, where CMB-polarization information is well measured.  If we were to try and use galaxy-survey data by itself, we would notice increased uncertainty in the tilt $n_s$.

\section{Conclusions} 
\label{sec:conclude}
In this paper, we have discussed the possibility that patchy reionization may leave an imprint in the distribution of galaxies through its effect on the collapse and cooling of baryons.  We considered a simple {\em ansatz} linking galaxy number density to the ionization fraction and used a halo-model approach to calculate the imprint of inhomogenous ionization on the galaxy power spectrum.  We then applied a Fisher-matrix approach to place constraints on the effect of this imprint.  

Our calculation shows that detecting the bubble imprint through large galaxy surveys is potentially feasible, but highly dependent upon the details of reionization.  We have shown that, for a detection to be possible with upcoming experiments, bubbles must be large ($r_{\rm{bub}}>10\,\rm{Mpc}$) and the feedback moderately strong ($\epsilon_b>0.1$).  This suggests that the most reasonable region of parameter space ($r_{\rm{bub}}<10\,\rm{Mpc}$, $\epsilon_b<0.1$) will not be detected with currently proposed galaxy surveys at $z\le3$.  Potentially, a $z\approx6$ galaxy survey might give the additional leverage needed for a concrete detection.

Beyond the possibility of detection, we have considered the effect of the bubble imprint on constraining dark-energy parameters.  We find that the distinctive nature of the baryon oscillations helps minimize any degeneracy arising.  Only if the characteristic bubble size is $\sim80\,\rm{Mpc}$ does the bubble imprint seriously impact our uncertainty in $w_0$.  In this case, the bubble power spectrum closely mimicks the cutoff of the density power spectrum.  This is a region of parameter space where the bubbles should be easily detected.  For more sensible values of ($r_{\rm{bub}}$,$\epsilon_b$), there is little or no impact on dark-energy constraints.  When the bubbles are not detectable, we find that ignoring them in the analysis of galaxy data does not introduce any significant biasing of the best-fitting parameters.  

Our approach has emphasised the use of a simple toy model to probe the effect of reionization on the distribution of galaxies.  If future galaxy surveys are able to make detections of this signal, it will be important to incorporate more detailed physics to better constrain the shape and amplitude of the bubble imprint.  With our present understanding of reionization, this seems premature.

Future galaxy surveys will greatly add to our knowledge of the distribution of galaxies and the nature of the dark energy.  If we are to extract maximum information from these surveys, we must tighten our understanding of the biasing of galaxy formation and the possible effect of reionization on early generations of galaxies.

This work was supported at Caltech in part by DoE DE-FG03-92-ER40701 and NASA NNG05GF69G.



\end{document}